# RAPIDMap: Rapid Multi-Agent Pipeline for Interpretable Disaster Mapping from Satellite and Street-view Imagery

**Yifan Yang [1], Lei Zou [1,*]**

[1] Department of Geography, Texas A&M University, College Station, TX, USA
* lzou@tamu.edu

## Introduction

As natural hazards become more frequent, rapid and reliable disaster mapping of impacted areas, damaged infrastructure, and affected populations is essential to enhance situational awareness, emergency resource allocation, and post-disaster recovery (Kirpalani, 2024; Yu et al., 2018; Kerle, 2024; Khan et al., 2023). Figure 1 shows a damage map of the early 2025 California wildfire using a commonly adopted WebGIS design. Each building is manually inspected and assigned a damage status (e.g., destroyed, major, minor, or unaffected), then georeferenced and visualized as a point feature in the WebGIS. However, such maps provide limited insight into the fine-grained extent and semantic characteristics of damage. Moreover, their generation typically requires substantial time and manual effort, as it involves field-based inspections, manual damage annotation, and multi-stage geospatial data integration.

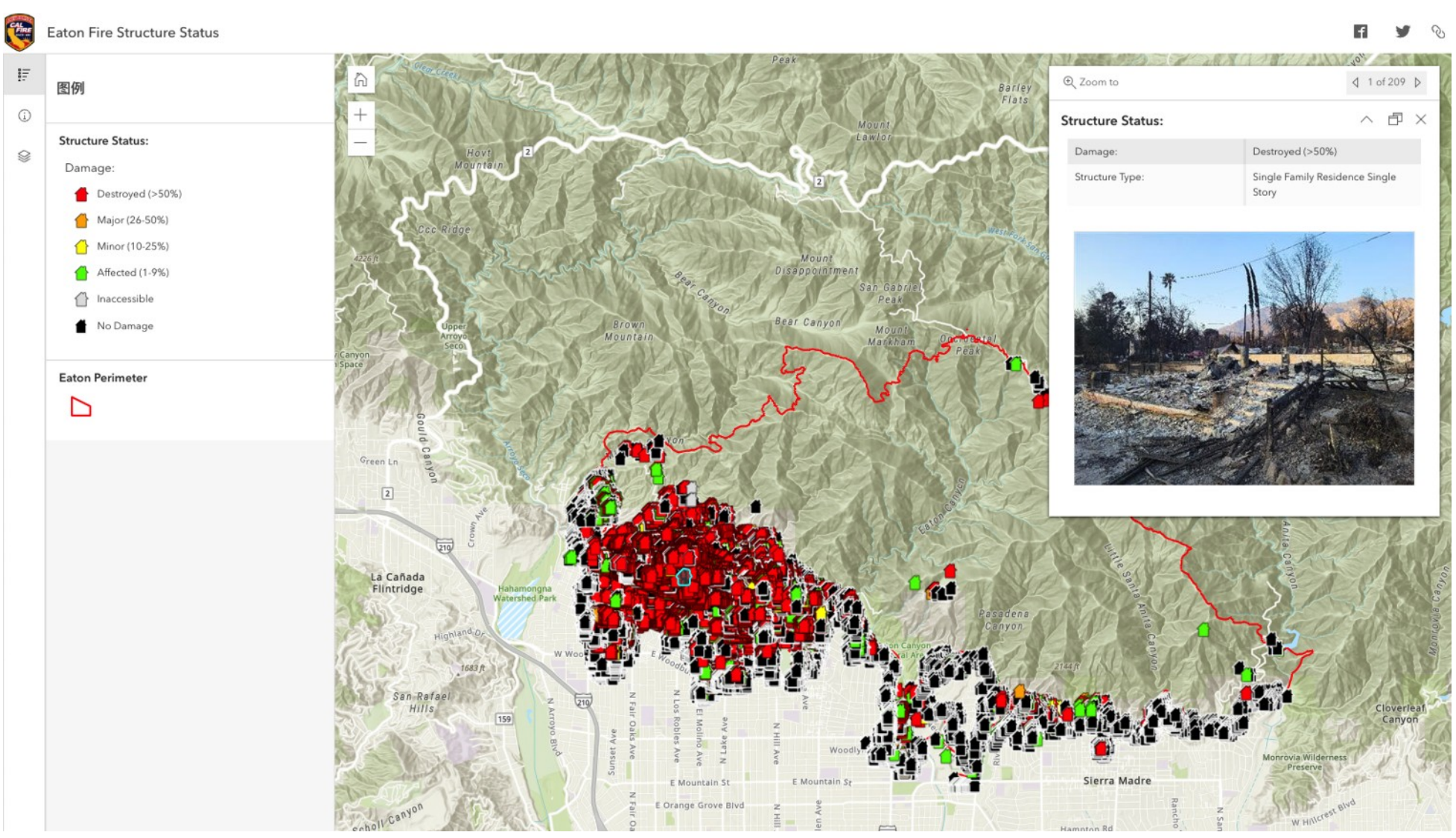

Figure 1: A WebGIS map of structural damage caused by the Early 2025 California Wildfire(https://lacounty.maps.arcgis.com/apps/instant/sidebar/index.html?appid=2209ecd140d8456291aeba7adbdfd69b)

In recent years, disaster mapping increasingly relies on multi-source geospatial data, including remote-sensing imagery from satellites, drones, and survey vehicles; crowdsourced data (e.g., location-based social media and mobile apps); and field surveys (e.g., site measurements and questionnaires). Advances in AI, particularly large-scale pre-trained models, enable more accurate and scalable extraction of disaster impacts from these heterogeneous data sources. However, leveraging pre-trained AI models for timely, spatially detailed disaster mapping remains challenging. First, many existing approaches rely on fine-tuning large pre-trained models with manually annotated datasets, but creating such datasets requires considerable time and labor, making it difficult to implement in time-sensitive disaster scenarios (Yang et al., 2025; Yang et al., 2026). Second, previous attempts trained models on datasets from a single disaster type or event, limiting generalization across regions and hazard types (Ma et al., 2025). Third, traditional disaster analysis often relies on single-modal observations, such as satellite imagery or street-view images alone, neglecting the complementary value of multi-source data in capturing both regional spatial context and ground-level damage characteristics (Chen et al., 2024; Lei et al., 2025; Chen et al., 2025; Ahn et al., 2025; Ma et al., 2025).

To address these challenges, this paper proposes **RAPIDMap**, a **R**apid multi-**A**gent **P**ipeline for **I**nterpretable **D**isaster **Map**ping from satellite and street-view imagery using zero-shot AI agents. The proposed framework provides an AI-driven interpretation layer that combines street-view and remote-sensing observations to generate disaster maps with rich, location-based semantic information. This study investigates whether zero-shot AI agents, when orchestrated within an end-to-end mapping pipeline, can enable rapid, accurate, and interpretable disaster mapping across diverse geographic contexts and disaster scenarios. This design directly addresses the three major challenges mentioned above: (1) it eliminates the need for manually annotated training data or model fine-tuning, enabling deployment within hours of a disaster; (2) the processing pipeline was validated using datasets covering multiple disaster categories, demonstrating its generalizability across different regions and disaster types; and (3) the pipeline explicitly integrates satellite and street-view imagery, combining broad spatial context with ground-level evidence of damage.

## Methodology

### *Multi-Agent Pipeline*

As shown in Figure 2, this study constructs a multi-agent disaster damage assessment and mapping pipeline composed of four core intelligent agents. The *Disaster Perception Agent (DPA)* serves as the entry layer, performing zero-shot recognition of data modality and disaster type from inputs and generating structured task-planning signals for downstream modules. The *Image Restoration Agent (IRA)* operates as a quality-control unit, diagnosing potential degradations in street-view and remote sensing imagery and executing constrained enhancement strategies when necessary to preserve disaster-relevant visual evidence. The *Damage Recognition Agent (DRA)* performs structured damage diagnosis across cross-view and cross-temporal settings, producing standardized severity classifications, object-level indicators, and confidence scores without task-specific fine-tuning. Finally, the *Disaster Mapping Agent (DMA)* functions as the spatial integration and visualization layer, aggregating structured damage outputs and projecting them into geospatial representations. It performs geo-referencing, cross-view alignment between remote sensing and street-view data, and generates map-based visualizations and GIS-ready structured outputs for downstream analysis and decision-making.

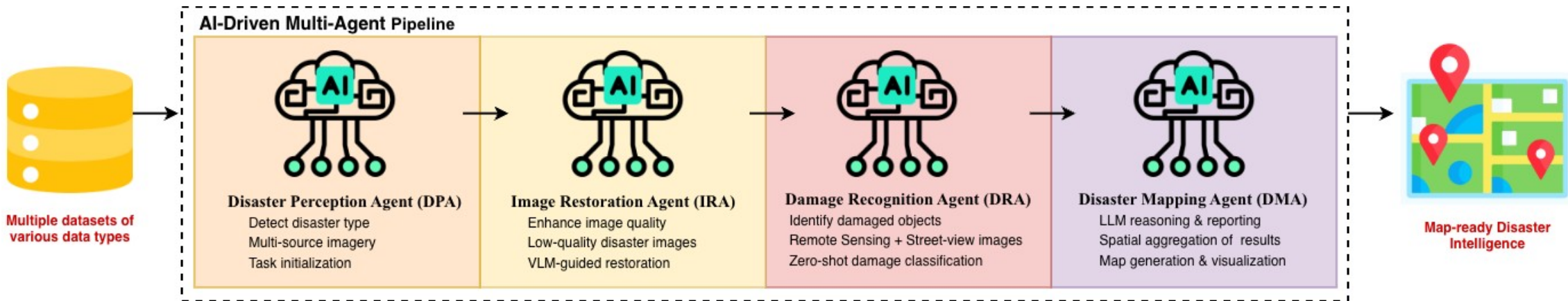

Figure 2: The proposed RAPIDMap multi-agent disaster assessment and mapping framework.

Each agent is instantiated using a suite of advanced multimodal and language models. For DPA, IRA, and DRA, we evaluated multiple model backbones, including GPT-5.1, GPT-5.1-mini, Gemini-2.5-flash, Gemini-2.5-Pro, and Gemini-3-Pro, enabling cross-model comparison and robustness analysis. For DMA, we introduced an additional reasoning and evaluation module powered by GPT-5.2, which generates structured, map-ready outputs and conducts a map-level assessment of disaster information.

### *Experimental Datasets*

Based on the complementary characteristics of disaster images and data availability, we selected representative disaster image datasets according to three considerations: cross-view, bi-temporal, and multi-hazard coverage. We constructed an experimental data system comprising three core types of disaster image combinations: (A) cross-view images (remote sensing & street-view), (B) bi-temporal street-view images (pre-disaster & post-disaster street-view), and (C) multi-hazard post-disaster street-view images (post-disaster street-view). The datasets cover typical disaster events such as hurricanes and wildfires, spanning multiple disaster areas in the United States, including California and Florida. Figure 3 shows the geographical distribution of the disasters included in this study, along with example images.

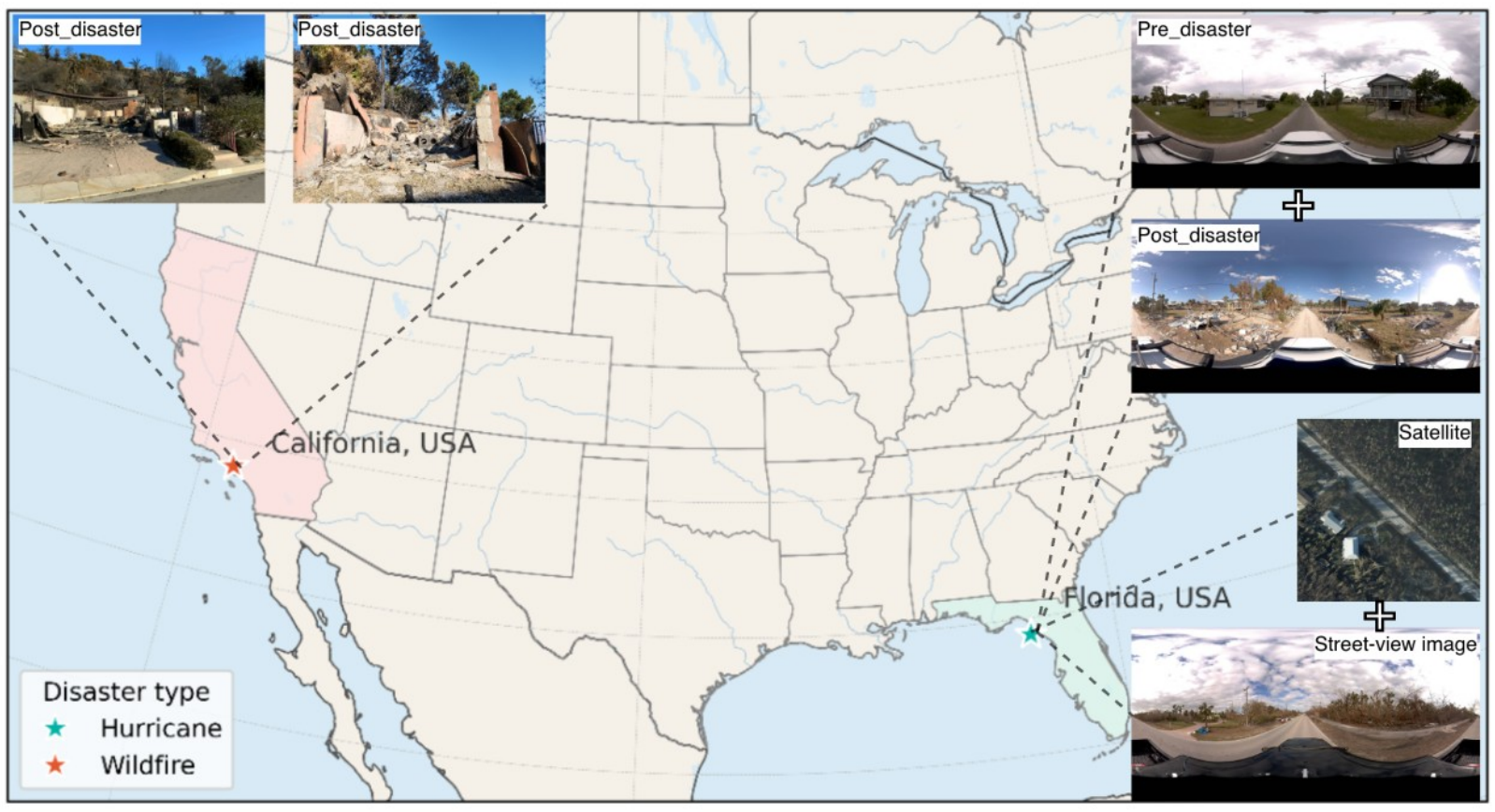

Figure 3: Geographic locations of the disasters included in this study.

Table 1 summarizes the datasets used and their roles in examining the performance of different AI agents in RAPIDMap.

Table 1: Types, composition, and characteristics of the disaster imagery datasets used in this study.

| Dataset | Data Type | Images | Disaster | Severity | Source | Associated Agent(s) |
|---|---|---|---|---|---|---|
| A | SVI+RSI pairs | 300 | Hurricane Ian, 2022 | 3 levels | CVDisaster (Li et al., 2025) | DPA, IRA, DRA, DMA |
| B | Bi-temporal SVI | 300 | Hurricane Milton, 2024 | 3 levels | Bi-temporal (Yang et al., 2025) | DPA, IRA, DRA, DMA |
| C1 | Post-disaster SVI | 188 | Drought (40), Earthquake (36), Flood (38), Ice Storm (44), and Wildfire (30) | N/A | Incidents Dataset (Weber et al., 2020) | DPA |
| C2 | Post-disaster SVI | 295 | Wildfire | 5 levels | LA DINS (2025) | DPA, DRA |

## Results

### *Disaster Perception Agent (DPA)*

We evaluated DPA on a disaster type classification task, where the agent assigns each case (i.e., input image) to one of seven categories (drought, earthquake, flood, hurricane, ice storm, wildfire, and others). We compared three Large Language Model (LLM) backbones and reported per-class precision, recall, F1-score, and overall accuracy (Table 2). Overall, all models performed strongly on this task, achieving overall accuracies between 0.86 and 0.92. Recognizing earthquake-related images was the most challenging category, with F1-scores ranging from 0.65 to 0.71.

Table 2: Performance comparison of Disaster Perception Agent (DPA) in multi-disaster type classification.

| Model | | Drought | Earthquake | Flood | Hurricane | Ice storm | Wildfire |
|---|---|---|---|---|---|---|---|
| Precision ↑ | GPT-5.1-mini | 0.95 | 0.91 | 0.90 | 0.98 | 1.00 | 0.99 |
| | GPT-5.1 | 0.97 | 0.95 | 0.92 | 0.97 | 0.97 | 0.99 |
| | Gemini-2.5-flash | 0.95 | 0.95 | 0.90 | 0.91 | 1.00 | 0.99 |
| Recall ↑ | GPT-5.1-mini | 0.93 | 0.58 | 0.97 | 0.99 | 0.86 | 0.98 |
| | GPT-5.1 | 0.85 | 0.56 | 0.92 | 1.00 | 0.70 | 0.97 |
| | Gemini-2.5-flash | 0.93 | 0.50 | 0.92 | 1.00 | 0.50 | 0.99 |
| F1-score ↑ | GPT-5.1-mini | 0.94 | 0.71 | 0.94 | 0.99 | 0.93 | 0.98 |
| | GPT-5.1 | 0.91 | 0.70 | 0.92 | 0.99 | 0.82 | 0.98 |
| | Gemini-2.5-flash | 0.94 | 0.65 | 0.91 | 0.95 | 0.67 | 0.99 |
| Overall Accuracy ↑ | GPT-5.1-mini | 0.92 | | | | | |
| | GPT-5.1 | 0.88 | | | | | |
| | Gemini-2.5-flash | 0.86 | | | | | |

***Image Restoration Agent (IRA)***

Table 3 summarizes the performance of different pre-trained models on the IRA task. We used a composite image quality score (Q) that integrates normalized contrast (C), sharpness (S), and an NIQE proxy (N) to reflect restoration effectiveness. We calculate it as $Q=0.4C+0.4S-0.2N$, where higher values indicate better visual quality. On satellite images, the baseline and planner-guided tool chains achieve the largest gains (Q: 0.62→0.73/0.71), while the Gemini image-only model performs worse because of limited structural cues. In contrast, on street-view images, the Gemini model performs best (0.75→0.79), surpassing both the baseline and planner methods.

Table 3: Performance comparison of Image Restoration Agent (IRA) across different input modalities and disaster types.

| Category | Total amount | Disaster type | Image type | Number of restoration | $Q_{original}$ | $Q_{baseline}$ | $Q_{planner}$ | $Q_{gemini}$ |
|---|---|---|---|---|---|---|---|---|
| A (Post-disaster SVI + RSI pairs) | 150 | Hurricane | Satellite | 141 | 0.62 | 0.73 | 0.71 | 0.69 |
| | | Hurricane | SVI | 22 | 0.75 | 0.78 | 0.76 | 0.79 |
| B (Pre/Post SVI pairs) | 150 | Hurricane | SVI | 4 | 0.76 | 0.78 | 0.78 | 0.79 |
| C (Post-disaster SVI) | 471 | Wildfire | SVI | 14 | 0.61 | 0.67 | 0.67 | 0.57 |
| | | Tropical | SVI | 2 | 0.51 | 0.55 | 0.62 | 0.60 |
| | | Ice storm | SVI | 4 | 0.73 | 0.74 | 0.75 | 0.67 |
| | | Flooded | SVI | 2 | 0.68 | 0.73 | 0.77 | 0.71 |
| | | Earthquake | SVI | 2 | 0.71 | 0.79 | 0.80 | 0.77 |
| | | Drought | SVI | 2 | 0.66 | 0.75 | 0.78 | 0.61 |
| | 295 | Wildfire | SVI | 3 | 0.70 | 0.77 | 0.78 | 0.78 |

Figure 4 provides a qualitative comparison of restoration results across different enhancement strategies for both SVI and remote sensing RSI samples. The visualization illustrates how baseline enhancement, planner-guided tool chains, and Gemini-based image-only optimization affect illumination, contrast, structural clarity, and disaster-relevant details under heterogeneous degradation conditions.

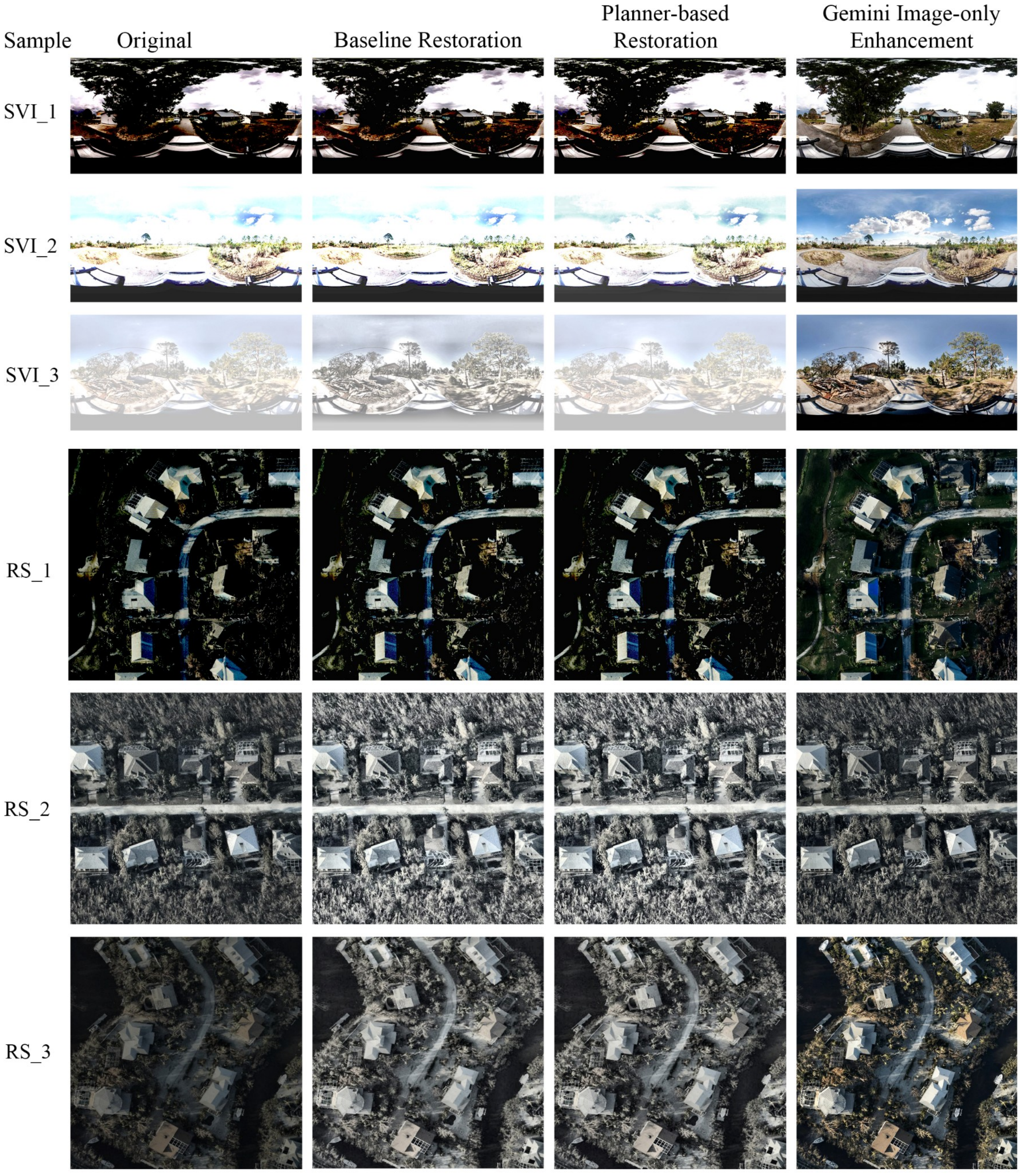

Figure 4: Visual comparison of restoration outputs produced by the Image Restoration Agent across baseline enhancement, planner-based restoration, and Gemini image-only optimization for SVI and RSI samples.

***Damage Recognition Agent (DRA)***

Table 4 compares the DRA's overall accuracy and Normalized Cross-Severity Error (NCSE) across three datasets, each containing damage-severity labels for hurricanes or wildfires. Overall accuracy is computed as the proportion of test images whose predicted damage-severity level exactly matches the ground-truth severity label over all evaluated images in each dataset (Table 1). NCSE additionally weights misclassifications by their distance from the true severity level and normalizes the result to [0, 1], so that confusing distant severity levels is penalized more heavily than confusing adjacent ones.Overall, performance is dataset-dependent, but in all three cases, the model with the highest accuracy also attains the lowest NCSE, indicating consistent gains in both correctness and severity-aware reliability.

Table 4: Performance comparison of the Damage Recognition Agent in multi-disaster severity levels.

| Model | | Hurricane (Dataset A) | Hurricane (Dataset B) | Wildfire (Dataset C) |
|---|---|---|---|---|
| Overall Accuracy ↑ | GPT-5-mini | 0.387 | 0.503 | 0.573 |
| | GPT-5.1 | 0.573 | **0.591** | 0.570 |
| | Gemini-2.5-flash | 0.360 | 0.470 | 0.559 |
| | Gemini-2.5-Pro | 0.380 | 0.447 | **0.590** |
| | Gemini-3-Pro | **0.627** | 0.493 | 0.442 |
| Normalized Cross-Severity Error ↓ | GPT-5-mini | 0.307 | 0.248 | 0.165 |
| | GPT-5.1 | 0.213 | **0.218** | 0.173 |
| | Gemini-2.5-flash | 0.363 | 0.299 | 0.179 |
| | Gemini-2.5-Pro | 0.373 | 0.327 | **0.162** |
| | Gemini-3-Pro | **0.190** | 0.291 | 0.210 |

The confusion matrices in Figure 5 further illustrate the error patterns of these best-performing models.

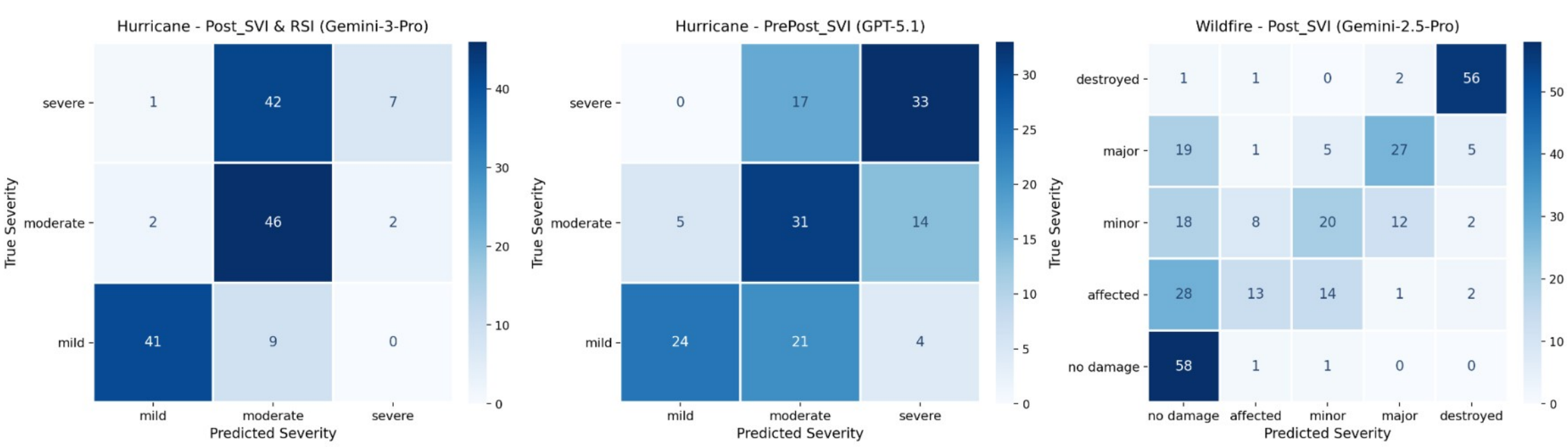

Figure 5: Confusion matrices of the best-performing model across three datasets.

Table 5 systematically compares the damage classification performance of various multimodal large models on two types of tasks: (1) Cross-view post-disaster data (Dataset A): joint inference using satellite images and street view images; (2) Dual-temporal street view data (Dataset B): comparative inference using pre-disaster images from 2023 and post-disaster images from 2024.

Table 5: Quantitative performance of different models on disaster damage recognition across cross-view hurricane imagery (Dataset A) and bi-temporal street-view imagery (Dataset B).

| Category | Model | F1-score | Recall | Precision | Accuracy | Damaged Building | Debris | Downed Power Line | Fallen Tree | Flooded Area |
|---|---|---|---|---|---|---|---|---|---|---|
| Hurricane Dataset A (Post-disaster street-view and remote-sensing image pairs) | Gemini-3-Pro | **0.77** | 0.78 | **0.80** | **0.93** | **0.94** | 0.89 | **1.00** | **0.92** | 0.88 |
| | Gemini-2.5-Pro | 0.57 | **0.91** | 0.50 | 0.72 | 0.57 | 0.77 | 0.67 | 0.63 | **0.95** |
| | Gemini-2.5-flash | 0.55 | 0.81 | 0.54 | 0.75 | 0.53 | 0.76 | 0.83 | 0.67 | 0.94 |
| | GPT-5.1 | 0.54 | 0.58 | 0.64 | 0.84 | 0.74 | **0.84** | 0.97 | 0.74 | 0.91 |
| | GPT-5-mini | 0.50 | 0.58 | 0.46 | 0.79 | 0.75 | 0.81 | 0.92 | 0.60 | 0.90 |
| Hurricane Dataset B (Pre- and post-disaster street-view image pairs) | Gemini-3-Pro | 0.39 | 0.36 | 0.51 | 0.85 | 0.74 | 0.80 | 0.99 | 0.72 | 0.92 |
| | Gemini-2.5-Pro | 0.43 | 0.47 | 0.40 | 0.79 | 0.72 | 0.79 | 0.77 | 0.69 | 0.79 |
| | Gemini-2.5-flash | 0.43 | 0.44 | 0.43 | 0.83 | 0.78 | 0.78 | 0.93 | 0.69 | 0.88 |
| | GPT-5.1 | **0.55** | **0.54** | **0.57** | **0.96** | **0.97** | **0.92** | **1.00** | **0.91** | **0.99** |
| | GPT-5-mini | 0.47 | 0.49 | 0.45 | 0.86 | 0.78 | 0.82 | 0.97 | 0.74 | 0.79 |

## *Disaster Mapping Agent (DMA)*

Figure 6 illustrates the inference quality assessment results of different large-scale language models in generating disaster reports on two different types of datasets. The assessment was conducted by comparing an automatic LLM-generated assessment with a manual assessment across four dimensions: factual consistency, plausibility, completeness of information, and actionability of recovery recommendations, with an overall score calculated.

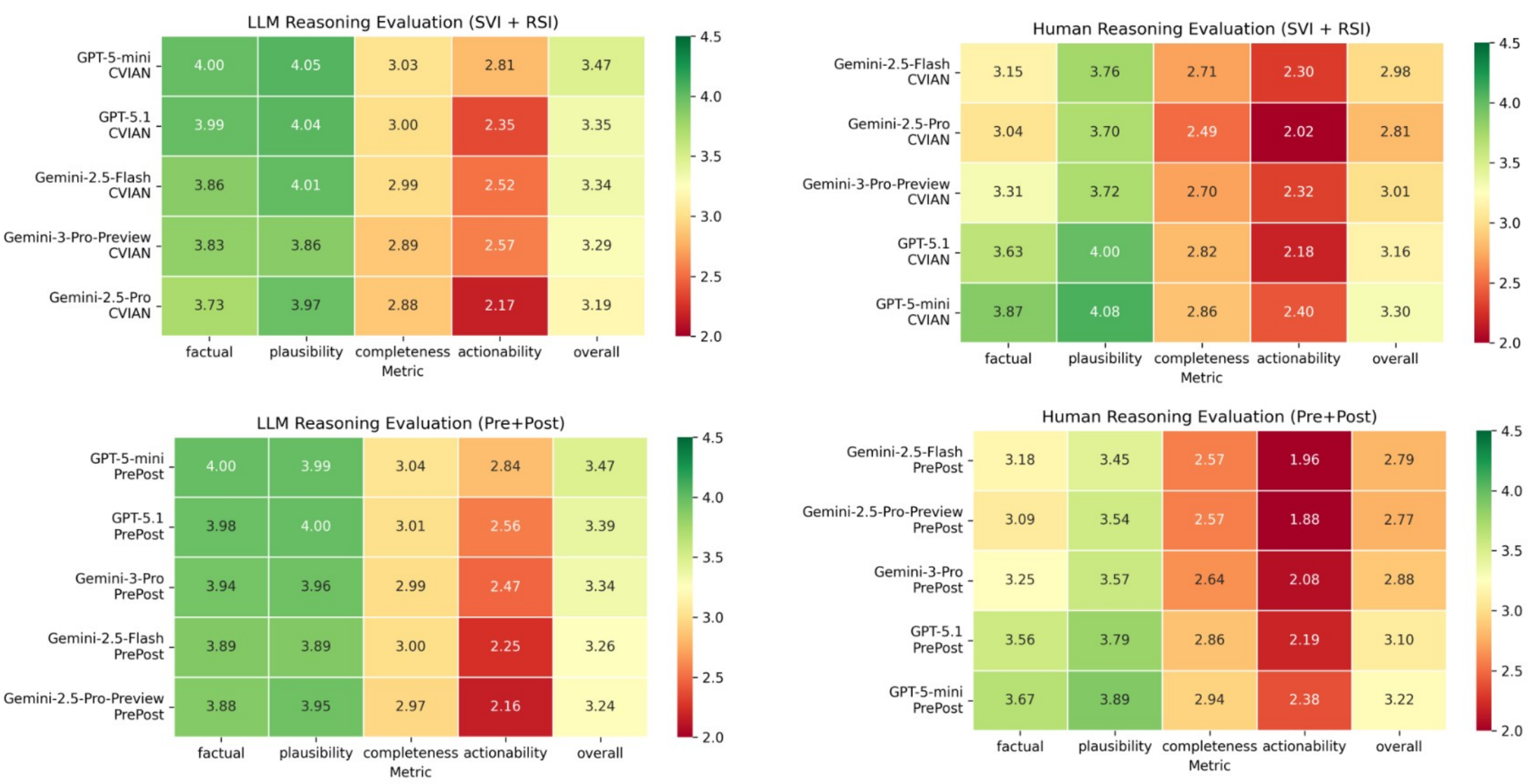

Figure 6: LLM and Human Evaluation of Multimodal Disaster Reasoning.

Unlike traditional disaster maps that primarily visualize the spatial distribution of damage, our framework enables AI-generated disaster intelligence for each mapped location, including disaster type identification, damage severity estimation, object recognition, confidence scores, and recovery recommendations (Figure 7).

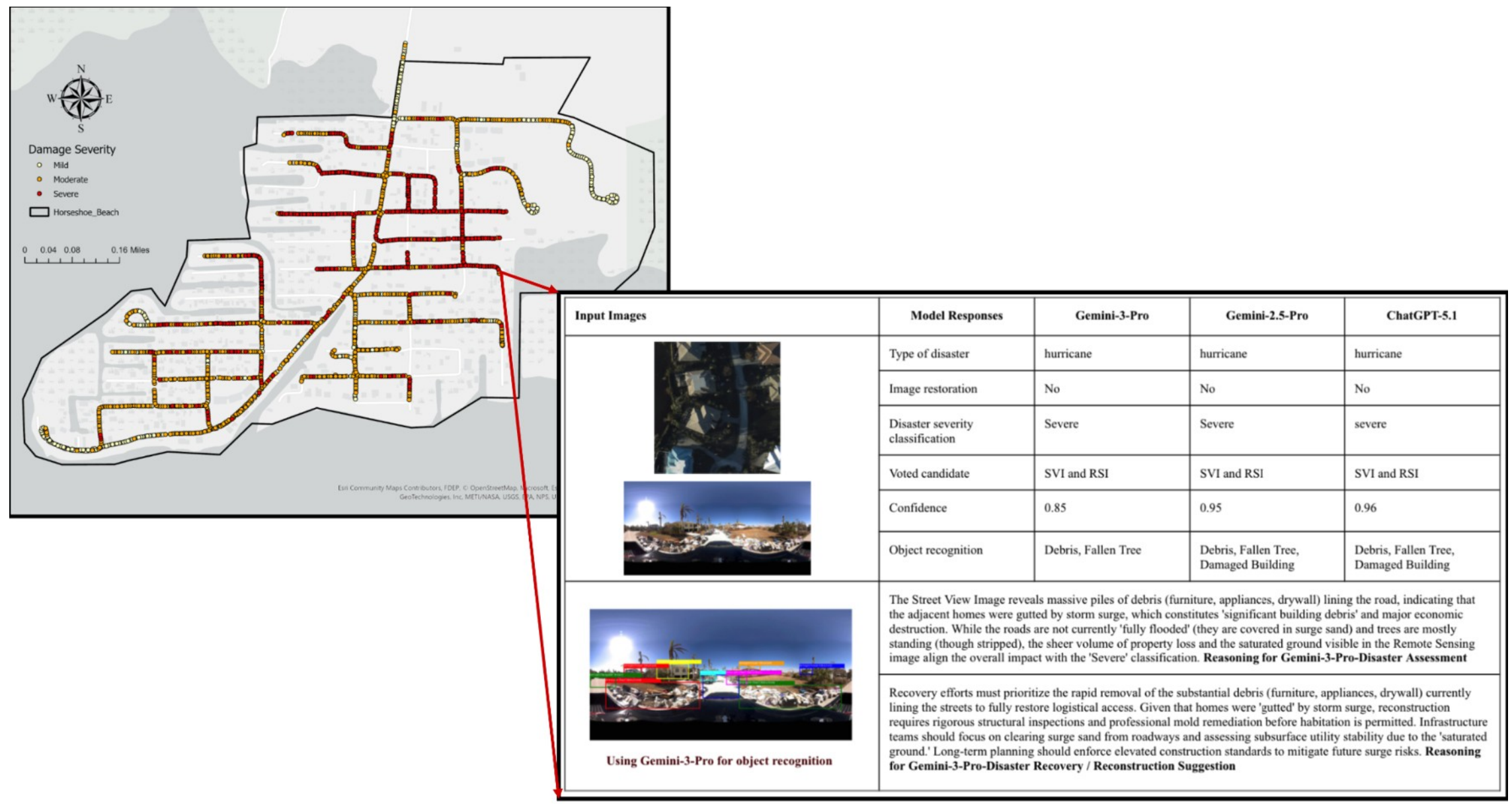

| Input Images | Model Responses | Gemini-3-Pro | Gemini-2.5-Pro | ChatGPT-5.1 |
|---|---|---|---|---|
| | Type of disaster | hurricane | hurricane | hurricane |
| | Image restoration | No | No | No |
| | Disaster severity classification | Severe | Severe | severe |
| | Voted candidate | SVI and RSI | SVI and RSI | SVI and RSI |
| | Confidence | 0.85 | 0.95 | 0.96 |
| | Object recognition | Debris, Fallen Tree | Debris, Fallen Tree, Damaged Building | Debris, Fallen Tree, Damaged Building |
| **Using Gemini-3-Pro for object recognition** | The Street View Image reveals massive piles of debris (furniture, appliances, drywall) lining the road, indicating that the adjacent homes were gutted by storm surge, which constitutes 'significant building debris' and major economic destruction. While the roads are not currently 'fully flooded' (they are covered in surge sand) and trees are mostly standing (though stripped), the sheer volume of property loss and the saturated ground visible in the Remote Sensing image align the overall impact with the 'Severe' classification. **Reasoning for Gemini-3-Pro-Disaster Assessment** | | | |
| | Recovery efforts must prioritize the rapid removal of the substantial debris (furniture, appliances, drywall) currently lining the streets to fully restore logistical access. Given that homes were 'gutted' by storm surge, reconstruction requires rigorous structural inspections and professional mold remediation before habitation is permitted. Infrastructure teams should focus on clearing surge sand from roadways and assessing subsurface utility stability due to the 'saturated ground.' Long-term planning should enforce elevated construction standards to mitigate future surge risks. **Reasoning for Gemini-3-Pro-Disaster Recovery / Reconstruction Suggestion** | | | |

Figure 7: Example of spatially explicit disaster mapping results generated by the proposed pipeline.

## Conclusion

This paper presents RAPIDMap, a multi-agent pipeline for zero-shot disaster damage assessment and mapping using multimodal imagery. By integrating remote sensing and street-view data, the framework enables automated disaster perception, damage recognition, and reasoning to generate structured disaster intelligence. The results demonstrate the potential of AI-driven approaches to enhance disaster mapping and support timely decision-making for emergency response and recovery.

## Acknowledgements:

This research is based on work supported by the National Academies of Sciences, Engineering, and Medicine (NASEM) under the Gulf Research Program (SCON-10000653, SCON-10001536) and the U.S. National Science Foundation (2318206).

## References

Ahn, K., Han, S., Park, S., Kim, J., Park, S., & Cha, M. (2025, April). Generalizable disaster damage assessment via change detection with vision foundation model. In *Proceedings of the AAAI Conference on Artificial Intelligence* (Vol. 39, No. 27, pp. 27784-27792).

Chen, H., Song, J., Dietrich, O., Broni-Bediako, C., Xuan, W., Wang, J., ... & Yokoya, N. (2025). BRIGHT: A globally distributed multimodal building damage assessment dataset with very-high-resolution for all-weather disaster response. *Earth System Science Data Discussions*, *2025*, 1-51.

Chen, Z., Shamsabadi, E. A., Jiang, S., Shen, L., & Dias-da-Costa, D. (2024). Integration of large vision language models for efficient post-disaster damage assessment and reporting. *arXiv preprint arXiv:2411.01511*.

Kerle, N. (2024). Disasters: Risk assessment, management, and post-disaster studies using remote sensing. In *Remote Sensing Handbook, Volume VI* (pp. 153-198). CRC Press.

Khan, S. M., Shafi, I., Butt, W. H., Diez, I. D. L. T., Flores, M. A. L., Galán, J. C., & Ashraf, I. (2023). A systematic review of disaster management systems: approaches, challenges, and future directions. *Land*, *12*(8), 1514.

Kirpalani, C. (2024). Technology-driven approaches to enhance disaster response and recovery. *Geospatial Technology for Natural Resource Management*, 25-81.

Lei, Z., Dong, Y., Li, W., Ding, R., Wang, Q. R., & Li, J. (2025, July). Harnessing large language models for disaster management: A survey. In Findings of the Association for Computational Linguistics: ACL 2025 (pp. 14528-14551).

Li, H., Deuser, F., Yin, W., Luo, X., Walther, P., Mai, G., ... & Werner, M. (2025). Cross-view geolocalization and disaster mapping with street-view and VHR satellite imagery: A case study of Hurricane IAN. *ISPRS Journal of Photogrammetry and Remote Sensing*, *220*, 841-854.

Ma, Z., Li, L., Li, J., Hua, W., Liu, J., Feng, Q., & Miura, Y. (2025). A Multimodal, Multilingual, and Multidimensional Pipeline for Fine-grained Crowdsourcing Earthquake Damage Evaluation. arXiv preprint arXiv:2506.03360.

Weber, E., Marzo, N., Papadopoulos, D. P., Biswas, A., Lapedriza, A., Ofli, F., ... & Torralba, A. (2020). Detecting natural disasters, damage, and incidents in the wild. In *Computer Vision – ECCV 2020* (pp. 331-350). Springer.

Yang, Y., Zou, L., Gong, W., Fu, K., Li, Z., Wang, S., ... & Tian, H. (2026). DamageArbiter: A CLIP-Enhanced Multimodal Arbitration Framework for Hurricane Damage Assessment from Street-View Imagery. arXiv preprint arXiv:2603.14837.

Yang, Y., Zou, L., Zhou, B., Li, D., Lin, B., Abedin, J., & Yang, M. (2025). Hyperlocal disaster damage assessment using bi-temporal street-view imagery and pre-trained vision models. *Computers, Environment and Urban Systems*, *121*, 102335.

Yu, M., Yang, C., & Li, Y. (2018). Big data in natural disaster management: a review. *Geosciences*, *8*(5), 165.